\documentclass[aip,amsmath,amssymb,reprint]{revtex4-2}
\usepackage[normalem]{ulem}
\usepackage[utf8]{inputenc}
\usepackage{amsmath}
\usepackage{amssymb}
\usepackage{figsize}
\usepackage{graphicx}
\usepackage[dvipsnames]{xcolor}
\usepackage{bm}
\usepackage{booktabs} 
\usepackage{comment}
\usepackage{subfigure}

\newcommand{\dd}{\mathrm{d}} 
\newcommand{\f}[2]{\mathchoice%
			{\dfrac{#1}{#2}}
	    	{\dfrac{#1}{#2}}
			{\frac{#1}{#2}}
			{\frac{#1}{#2}}}
\newcommand{\ddf}[3][]{\ifthenelse{\equal{#1}{}}{\ensuremath{\f{\dd#2}{\dd#3}}}
{\ensuremath{\f{\dd^{#1}#2}{\dd{#3}^{#1}}}}}
\newcommand{\Dp}[3][]{\ifthenelse{\equal{#1}{}}{\ensuremath{\f{\partial#2}{\partial#3}}}
  {\ensuremath{\f{\partial^{#1}#2}{\partial{#3}^{#1}}}}}

\begin{document}

\title{Correcting systematic parametrization errors in  underdamped Langevin models of molecular dynamics trajectories}

\author{David Daniel Girardier}
\email{david.girardier@sorbonne-universite.fr}
\affiliation{Sorbonne Université, Musée National d’Histoire Naturelle, UMR CNRS 7590, Institut de
Minéralogie, de Physique des Materiaux et de Cosmochimie, IMPMC, F-75005 Paris, France}
\author{Hadrien Vroylandt}
\affiliation{Université Caen Normandie, ENSICAEN, CNRS, Normandie Univ, GREYC UMR 6072, F-14000 Caen, France}
\author{Sara Bonella}
\affiliation{Centre Européen de Calcul Atomique et Moléculaire (CECAM), Ecole Polytechnique Fédérale
de Lausanne, Lausanne 1015, Switzerland}
\author{Fabio Pietrucci}
\affiliation{Sorbonne Université, Musée National d’Histoire Naturelle, UMR CNRS 7590, Institut de
Minéralogie, de Physique des Materiaux et de Cosmochimie, IMPMC, F-75005 Paris, France}


\begin{abstract}

Since Kramers' pioneering work in 1940, significant efforts have been devoted to studying Langevin equations applied to physical and chemical reactions projected onto few collective variables, with particular focus on the inference of their parameters. While the inference for overdamped Langevin equations is well-established and widely applied, a notable gap remains in the literature for underdamped Langevin equation. This gap arises from the challenge of accessing velocities solely through finite differences of positions, resulting in spurious correlations. In this letter, we propose an analytical correction for these correlations, specifically designed for a likelihood-maximization algorithm that exploits short, non-ergodic trajectories that can be obtained at reasonable numerical cost. The accuracy and robustness of our approach are tested 
on a benchmark case and a realistic system. 
This work paves the way for applying generalized Langevin equation inference to  activated phenomena, such as chemical reactions, in several scientific domains.
\end{abstract}


\maketitle





In addition to providing ergodic canonical sampling for atomistic simulations of condensed phase systems, Langevin equations are used by a broad scientific community to describe the time evolution of generic observables. 
The flexibility of the equations' structure, accommodating both deterministic and Markovian or non-Markovian random forces, enables them to model successfully phenomena as diverse as phase transitions~\cite{moritz17}, climate-change effects on ocean currents~\cite{ditlevsen23}, and clocks~\cite{milburn20}.

In particular, many complex, high-dimensional physical and chemical processes can be effectively studied through the lens of Langevin dynamics when projected onto one (or a few) collective variable (CV), $q$ that tracks their progress~\cite{Zwanzig01}. Depending on the problem, different types of Langevin equations should be adopted. In this work, we focus on the underdamped, Markovian regime. In this case, the evolution equations are
%
\begin{equation}\label{eq:ULE}
    \begin{cases}
        \dot{q}=& v \\
\dot{v} =& -\f{1}{m}\f{dF}{dq} - \gamma v + \tilde{\sigma}\, \eta(t)
    \end{cases}
\end{equation}
where  $m$ is the effective CV mass, $F(q)$ the free energy landscape (or potential of mean force), $\gamma(q)$ is the friction, $\tilde{\sigma}$ is the amplitude of the random force, and $\eta(t)$ a time-uncorrelated Gaussian white noise. If $\tilde{\sigma}=\sqrt{2\frac{k_BT}{m}\gamma}$, 
fluctuation dissipation is enforced.
In general, 
$F(q)$ and $\gamma(q)$
are not available analytically and must be determined starting from observational time series of the CV and (in principle of) its velocity,
based, e.g., on experiments or 
molecular dynamics trajectories.

Furthermore, the stochastic equation is typically not employed in the differential form above, but 
via a numerical integrator with a finite time step~\cite{Vanden2006} $\tau$, that establishes the time resolution adopted to analyze the data. Such resolution is a key parameter:
depending on the degree of temporal coarse-graining, different types of Langevin equations act as suitable approximations~\cite{girardier2023}. 
In the case of overdamped models,
statistical inference approaches based on
likelihood maximization currently allow to reconstruct the potential of mean force, the diffusion coefficient (related to the friction) and the kinetic rate in a range of overdamped processes~\cite{Hummer05,Micheletti08,Zhang11,Crommelin11,Crommelin12,Baldovin19,Sicard21,Donati22,palacio2022}.
We remark that long, ergodic trajectories that accurately sample the probability landscape of the CV are often unavailable, so it is important to develop numerical schemes able to exploit also short, unbiased, non-ergodic trajectories~\cite{palacio2022}.

The parametrization of underdamped models is presently considerably less developed and understood. These models are, however, needed to describe important phenomena, such as migrating cells dynamics~\cite{bruckner20} and biomolecular processes~\cite{Schaudinnus15}, that display inertial effects at fine-grained time resolution, so progress in this direction is highly desirable. Unfortunately, model inference for the underdamped case is significantly more complex than for the overdamped. 
CV positions and velocities appear on equal footings in  Eq.~\ref{eq:ULE}. However, often in observational time-series  only coordinates are directly accessible, while  velocities need to be estimated via finite differences schemes~\footnote{Moreover the 
equations of motion leading to the observed trajectory $q(t)$ are not necessarily of Langevin type
(think to an experiment, or to high-dimensional MD projected on a CV), so that Eq.~\ref{eq:ULE} is just a putative model of the original dynamics and it is meaningless to speak about `exact Langevin velocities'.
We also remark that, even remaining in the context of purely Langevin dynamics, Eq.~\ref{eq:ULE} can be numerically integrated using different algorithms~\cite{Vanden2006}, leading to different values of Langevin velocities.
}.  This introduces spurious correlations even at infinitely-small time resolution, spoiling the accuracy of current inference methods.  
In Ref.~\cite{ferretti20}
it was shown that serious inaccuracies on friction estimation (subsequently affecting free energy estimation in non ergodic calculations), arise when the propagator deployed in the likelihood scheme is of order lower than $\mathcal{O}(\tau ^{3/2})$.  Unfortunately, no general solution for this problem was proposed. The spurious velocity correlations hinder also use of alternative approaches.
For example, in Refs.~\cite{Schaudinnus15, Schaudinnus16,Lickert21} only positions are employed in the parametrization of the potential of mean force and the friction (and diffusion) is obtained via local averages. This procedure, however, is based on an ad hoc choice of an integrator of first order in time and therefore open to the problems highlighted in Ref. \cite{ferretti20}. 
A solution to the problems arising from finite-difference velocities was proposed
in Ref.\cite{bruckner20}, where unbiased estimators for the drift and noise amplitude in Eq.~\ref{eq:ULE} 
were derived. Even though this method represents a significant advance in the inference of realistic data, converging these estimators requires simulations lengths that might be unattainable in the presence of  free-energy barriers.  
The approaches mentioned above are not based on likelihood maximization.  More recently, we introduced a likelihood-based scheme in which the velocity issue was circumvented by inferring the friction beforehand through correlation functions within a free-energy well~\cite{girardier2023}. This approach has the advantage of simplicity but it does not account for the position-dependence of the friction and 
relies on approximately harmonic dynamics.


In this letter,  we propose a novel 
likelihood-maximization approach that overcomes the limitations of current inference schemes for underdamped Langevin models: 
requiring as only
input short, non ergodic trajectories, 
it eliminates the aforementioned systematic errors via an
original analytical correction scheme.
We provide both a priori 
and a posteriori evaluations of the quality of the inference that do not require prior knowledge on the system. 
We demonstrate the accuracy and robustness of the approach with respect to varying the time resolution of the Langevin model for benchmark and realistic systems. In principle, the approach can be extended to non-Markovian generalized Langevin equations.




The projection of high-dimensional Hamiltonian dynamics
onto one (or a few) CV(s) results in 
a generalized Langevin equation~\cite{Zwanzig01,vroylandt22FDT},
that simplifies to the Markovian Eq.~\ref{eq:ULE} for sufficiently coarse
temporal subsampling~\cite{girardier2023}. 
%
Starting from a given set of trajectories discretized as $M$ samples, 
an efficient strategy to parametrize Langevin models exploits the maximization of a likelihood function written as the product of short-time propagators:
%
 %
\begin{equation}\label{eq:lik}
    \mathcal{L}_\theta = \prod_{n=1}^{M-1} \mathcal{P}_\theta(q_{n+1},v_{n+1}|\,q_n,v_n)
  \end{equation}
where $\theta$ is the set of model parameter,
which in our case include friction and free-energy
landscapes. 
The mass is directly estimated
using the equipartition theorem or via a coordinate transformation ~\cite{jain2012equipartition,girardier2023}.

Following Refs.~\cite{Drozdov96,girardier2023} we approximate the propagator in Gaussian form
%
\begin{multline}\label{eq:prop}
    \mathcal{P}_\theta(q_{n+1},v_{n+1}|\,q_{n},v_{n}) = \\ 
    \frac{1}{\sqrt{4\pi |M|}} \exp\Big[ -\frac{M_{qq}}{2|M|}(v_{n+1} - \langle v\rangle)^2 \\
    - \frac{M_{vv}}{2|M|}(q_{n+1} - \langle q\rangle)^2 + \frac{M_{qv}}{|M|}(q_{n+1}-\langle q\rangle)(v_{n+1} - \langle v\rangle) \Big]
\end{multline}
where averages $\langle q \rangle,\langle v \rangle$ and (co)variances $M_{\alpha \beta}=\langle \alpha \beta\rangle-\langle \alpha \rangle\langle  \beta\rangle$ (elements of the matrix $M$) are defined in SI section A.

\begin{figure}%
    \centering
    {\includegraphics[width=8.5cm]{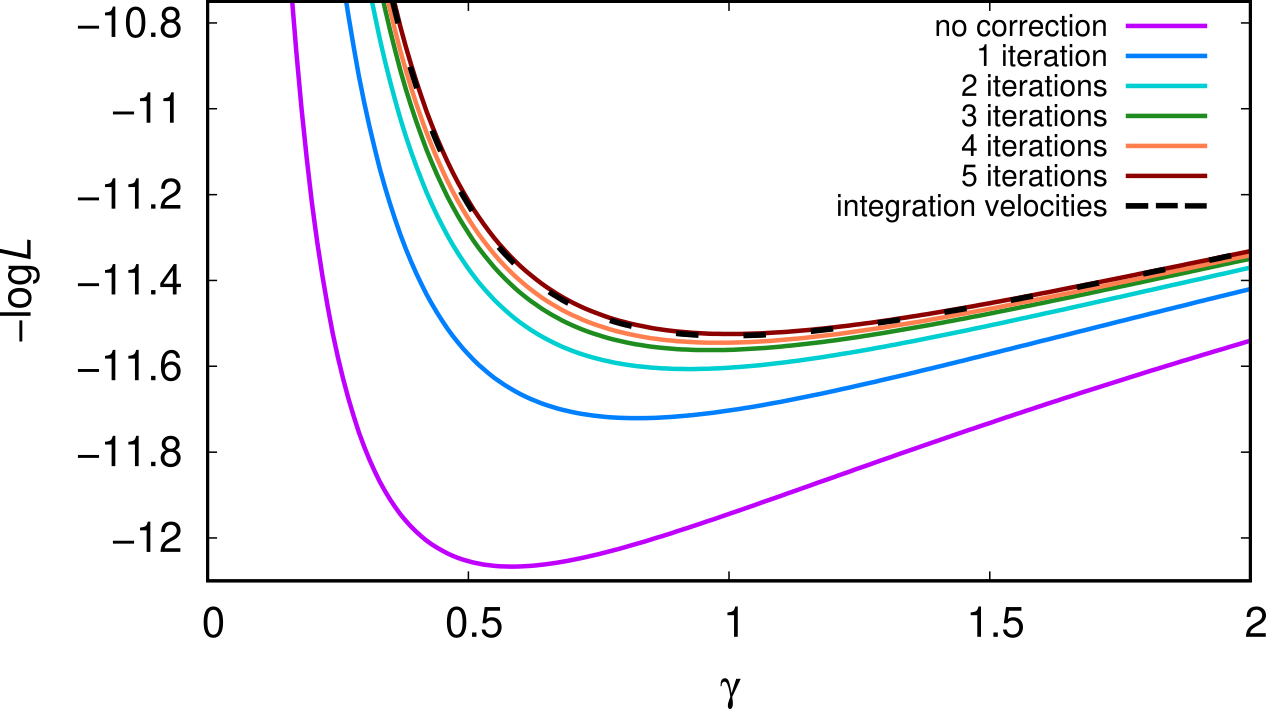} }%
    \caption{Corrected log-likelihood estimated from the finite-difference velocity using the first two equations in the set of Eq.~\ref{eq:vv}, as a function of the friction $\gamma$, for non-ergodic trajectories of a double-well system (with free energy and mass fixed to their optimal values), for different numbers of iterations of the correcting scheme. 
    The non-corrected curve is displayed for comparison.
    The dashed line represents the log-likelihood computed from the integration velocities.}
    \label{fig:LogL}%
\end{figure}

As proved in SI section B, the specific form of the matrix and averages adopted in this work corresponds, to order $\mathcal{O}(\tau ^3)$, to the evolution determined by the second order Vanden-Eijnden and Ciccotti (VEC) integrator for underdamped Langevin dynamics (eq. 21 in Ref.~\cite{Vanden2006}): 

%
\begin{align}\label{eq:VEC}    
    q_{n+1} =& q_{n}  + \tau \left( 1 -  \frac{\gamma\tau}{2}\right) v_{n} +  \frac{\tau^2}{2}\phi_{n}  + \frac{\tau}{2}\sigma  [G_{n} + \frac{1}{\sqrt{3}}\tilde{G}_{n}]\\
    v_{n+1} =&\left(1-\gamma\tau  +\frac{\gamma^2\tau^2}{2}\right) v_{n}  + \frac{\tau}{2}(\phi_{n}+\phi_{n+1}) + \sigma  G_{n} \nonumber\\
    & -\gamma\frac{\tau^2}{2}\phi_{n}  -\gamma \frac{\tau}{2}\sigma   [G_{n} + \frac{1}{\sqrt{3}}\tilde{G}_{n}] \nonumber 
\end{align}
with $\phi_n=-\frac{1}{m}\frac{dF(q_n)}{dq}$,  $\sigma = \tilde{\sigma}\sqrt{\tau}= \sqrt{2\frac{k_BT\gamma}{m}\tau}$. $G_n$ and $\tilde{G}_n$ are two independent Gaussian random numbers with zero mean and unit variance.
This correspondence can also be verified numerically. 
If exact positions and velocities of a VEC trajectory with integration time step $\tau$ are supplied to 
Eqs.~\ref{eq:lik},\ref{eq:prop}, the maximum likelihood value 
is obtained for the correct parameters of the underlying underdamped Langevin equation. This is illustrated in Fig.~\ref{fig:LogL}, that shows, for a double-well $F(q)$ landscape, different curves showing $-\log \mathcal{L}_\theta$ as a function of the friction. Let us focus, for the moment, on the dashed black curve in the figure, computed by feeding to the likelihood positions and velocities coming from a VEC trajectory: 
it yields an optimal $\gamma =1$, which is, indeed, the friction used to integrate the trajectory.

In typical applications, we need to estimate velocities starting from
the observed positions at a given time resolution $\tau$, resulting in a bias in the likelihood. This is illustrated via the purple curve in Fig.~\ref{fig:LogL}, 
obtained using finite-difference velocities. We now describe how this problem can be efficiently and accurately corrected. Let us consider the time-reversible~\cite{bruckner20} central finite difference estimator of the velocity $u_n = \f{ q_{n+1} - q_{n-1}}{2\tau}$, combined with the VEC integrator. 
In SI section C, it is shown that, up to order $\tau^{3/2}$, the difference between $u_n$ and the integration velocity $v_n$ is given by:
 \begin{equation}
 \label{eq:diffVel}
  v_{n} = u_{n}   -  \f{\sigma}{4}   \left[G_{n} + \f{1}{\sqrt{3}}\tilde{G}_{n} - G_{n-1}+\f{1}{\sqrt{3}}\tilde{G}_{n-1}\right] + \mathcal{O}(\tau^{3/2})
\end{equation}

Using this result, the properties of Gaussian white noises and their correlations with positions or velocities (SI section D), 
we can 
correct 
the averages appearing in the likelihood:
%
\begin{align} 
\label{eq:vv}
  &\langle v_n^2\rangle = \langle u_n^2\rangle + \frac{\sigma^2}{3} + \mathcal{O}(\tau^{3/2}) \nonumber \\
  & \langle v_n v_{n-1}\rangle = \langle u_n u_{n-1}\rangle + \frac{\sigma^2}{24}+ \mathcal{O}(\tau^{3/2}) \nonumber \\
  & \langle v_n q_n\rangle = \langle u_n q_n\rangle +\mathcal{O}(\tau^{3/2})\\
  &\langle v_{n} q_{n+1}\rangle = \langle u_{n} q_{n+1}\rangle + \mathcal{O}(\tau^{3/2}) \nonumber \\
  & \langle v_{n+1} q_n\rangle = \langle u_{n+1} q_n\rangle \nonumber
\end{align}
%
This enables to remove from the log-likelihood the spurious correlations that lead to substantial systematic errors for inferred free-energy and friction landscapes (see SI section E for details).
The corrections depend implicitly on the friction $\gamma$, one of the parameters that must be inferred, via $\sigma$. To circumvent this problem, we adopt an  expectation-maximization scheme:
 \begin{enumerate}
    \item likelihood optimization without correction
    \item correction of the likelihood with the previously obtained friction and new optimization 
    \item iterate until convergence
\end{enumerate}    
 The efficiency of the scheme is demonstrated in Fig.~\ref{fig:LogL}, where, after 5 iterations, the correction added to finite difference velocity likelihood (dark-red curve) recovers the likelihood computed using the integration velocity (dashed curve).

As a first full illustration of the validity and efficiency of the new algorithm, we infer
the parameters of the underdamped Langevin equation from one-dimensional underdamped Langevin trajectories in a double-well landscape with a $10\,k_BT$ barrier, $\gamma=5$~ps$^{-1}$ and unit mass, using an Euler-Maruyama integrator with a time step of $10^{-4}$~ps. Time resolutions $\tau \geq 10^{-3}$~ps are employed for analyses, to ensure that the choice of integrator does not affect the inference results.
Importantly, we consider relatively-short non-ergodic trajectories to train the model (Fig.~\ref{fig:Results}a), since in realistic barrier-crossing processes it is unfeasible to perform ergodic MD simulations.




In general, in order to determine whether the initial data can be accurately reproduced with an underdamped Langevin model, there are two characteristic timescales that define the {acceptable
bounds for inference. The lower bound is set by the 
correlation time of the noise for projected (non-Markovian) dynamics,
and the upper bound is determined by the velocity decorrelation time. The time-resolution adopted in the underdamped inference, $\tau$, has to be coarse-enough to lose non-Markovian effects, while providing sufficient resolution to capture the original velocity autocorrelation function (VACF). 
In our benchmark case, with Markovian dynamics, the lower limit is of no concern.  Fig.~\ref{fig:Results}c shows that, for our double well system, a time resolution $\tau\le 0.02$~ps
should be adequate.
\begin{figure*}[htbp]
    \centering
    {\includegraphics[width=1\linewidth]{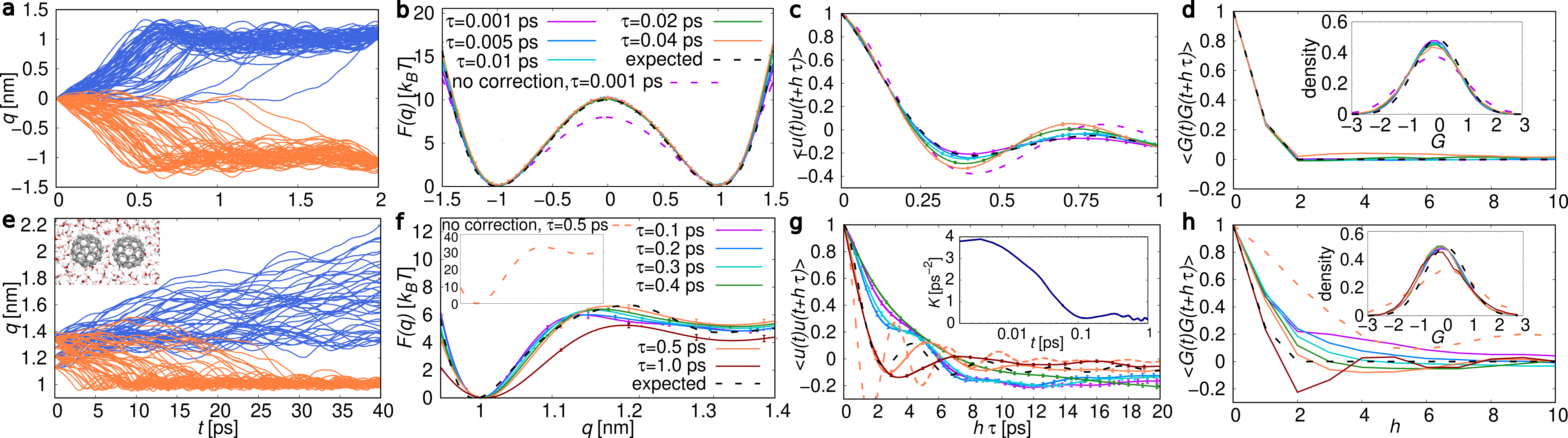} }
    \caption{Benchmarks for the new inference approach. Top panels: double well landscape; Bottom panels:  fullerenes in water. (a) Underdamped Langevin trajectories relaxing from the top of the barrier.
    (b) Free-energy profiles for underdamped models inferred at different time resolutions $\tau$ with the likelihood correction  (except the purple dashed curve, for comparison). (c) Velocity autocorrelation function of the original trajectories and of synthetic ones generated from the models using the Euler–Maruyama integrator. (d) Correlation analysis of the noise time series deduced from the trajectories in panel (a) by inverting the positional part of Vanden-Eijnden Ciccotti integrator. The inset shows the noise distribution.
    (e) Aimless shooting trajectories
    of a fullerene dimer in water (see inset) projected onto the distance between the C$_{60}$ centers of mass.  (f) Free-energy profiles at different time resolutions $\tau$ using the likelihood correction (inset: result without correction, with a $\approx 30~k_BT$ barrier). (g) Comparison between the velocity autocorrelation function of the original trajectories and of synthetic ones generated from the models using the Euler–Maruyama integrator (inset: memory kernel computed for the associated state of the dimer). (h) Correlation analysis of the noise time series deduced from the trajectories in (e) by inverting the positional part of the VEC integrator.The inset shows the noise distribution.}
    \label{fig:Results}%
\end{figure*}

Our iterative correction scheme is then applied to CV trajectories, for which the velocity is accessed through finite differences.
We then systematically compare the statistical properties of original trajectories with those of the underdamped models generated at different $\tau$ values. Results are shown in Fig.~\ref{fig:Results}, panels a-d. The sampling trajectories and results for the inferred free energy are presented in panel a and b of Fig.~\ref{fig:Results}. The free-energy landscape is accurately reproduced up to $\tau=0.04$~ps
 consistently with the VACF diagnostics in Fig.~\ref{fig:Results}c. As a further test, Fig.~\ref{fig:Results}d reports the distribution and time-correlation of the noise reconstructed from the trajectory {\sl via} the Langevin model, compared to the expected ideal behavior (derived in SI, Section F). Results for the friction are reported in SI section G where we show that it is accurately reproduced (error within 10\%) only up to $\tau = 0.01$~ps.

In summary, 
consistently with the diagnostics,
the most accurate models are obtained for 
$\tau$ in the range $0.001$–$0.01$~ps, even though the error remains moderate for higher $\tau$, highlighting the stability of our inference scheme with respect to the increase of time resolution.
Note that, without applying the correction scheme, the model is inaccurate even for an optimal choice $\tau=0.001$~ps. As shown by the purple dashed curve in Fig.~\ref{fig:Results}b, in this case,  
the barrier is underestimated by about 20\% and the friction by about 40\%, to be compared with the 1/3 relative error on the latter predicted in Ref.~\cite{ferretti20}.

We now focus on a realistic system, formed by two C$_{60}$ fullerene nanoparticles immersed in liquid water, modeled as a periodically-repeated cubic simulation box including 2398 water molecules (Fig.~\ref{fig:Results}e inset). We perform MD simulations at 300~K in the $NVT$ ensemble~\cite{Bussi07} , with the SPC water force field~\cite{Berendsen84} and
 OPLS-AA~\cite{Jorgensen96} for carbon using the GROMACS 2019.4 code~\cite{Berendsen95}, with a timestep of 1~fs. For computational details we refer to Ref.~\cite{girardier2023}.
We modified the masses of oxygen and carbon to 1 and 120~a.u., respectively to ensure, without affecting the free-energy landscape, a wider range of time resolutions for the accurate reconstruction of underdamped models (see SI section H for comparisons with unmodified masses).  

We analyze the dissociation/association process of the fullerene dimer, employing 500 short (40~ps) unbiased MD trajectories relaxing from the transition state ensemble, generated with the aimless shooting algorithm~\cite{Peters07,palacio2022}.
The high-dimensional trajectories are projected onto one CV, defined as the distance between the centers of mass of the two fullerenes (Fig.~\ref{fig:Results}e). Albeit sub-optimal, this CV captures the main features of the transition process. In particular, explicitly including water degrees of freedom does not improve the kinetic description~\cite{mouaffac23}.

As for the benchmark case, as a first step we determine a  time resolution $\tau$ that avoids memory effects while maintaining smoothness in the VACF. The memory kernel in Fig.~\ref{fig:Results}g,
computed using the Volterra method of Ref.~\cite{vroylandt22POSDEP} on a locally-equilibrated 10~ns trajectory of the dimer, indicates that $\tau\geq 0.1~\text{ps}$ is required to eliminate most non-Markovian effects, notwithstanding a surviving tail in the memory kernel. 
We make here the customary approximation of a position-independent kernel.

To ensure a good representation of the VACF (Fig.~\ref{fig:Results}g, black dashed line), the upper bound is set as $\tau< 1~\text{ps}$.
Thus, we conclude that the inference of underdamped Langevin models can be attempted for $0.1\leq\tau<1$~ps.
We apply our inference scheme in this range, obtaining $F(q)$ and $\gamma(q)$
by likelihood maximization, while estimating the mass via equipartition of kinetic energy at the finest time resolution $\tau = 10^{-3}$~ps, yielding 
$m=1.48\cdot 10^{3}$~ps$^2$~nm$^{-2}$.

As a first observation, we note that the inferred free-energy profile is in good agreement with the brute-force reference, particularly in the dimer region, except for 
the largest 
$\tau = 1$~ps. For $0.1\leq\tau\leq 0.5$~ps, the barrier height is consistent within $1\ k_BT$ across models, even though the transition-state region is best captured by the $\tau = 0.5$~ps model.
The importance of the likelihood correction can be appreciated from the inset in Fig.~\ref{fig:Results}f: the uncorrected model at the optimal $\tau$ 
yields a grossly inaccurate free-energy barrier ($\approx 30~k_BT$), consistent with the bad diagnostics in Fig.~\ref{fig:Results}g,h. Even for other values of $\tau$, the barrier is always overestimated. 
Comparison of the VACF for the different models with the one from the original MD simulation (Fig.~\ref{fig:Results}g) 
confirms that the time resolution $\tau = 0.5$~ps is the most suitable for accurately modeling the projected dynamics. The same resolution appears optimal also by testing the time-correlation of the noise inferred by interpreting the original MD trajectories as if they were generated by the Langevin models, compared with the expected analytical result
(Fig.~\ref{fig:Results}h).
We note that in the case of MD simulation data the interpretation of the noise correlation could be more subtle than for the benchmark one-dimensional system of Fig.~\ref{fig:Results}d, due to the long, non-exponential tail of the memory kernel.
The probability density of the noise, well described by a Gaussian of zero mean and $2/3$ variance for different $\tau$ values, points again to reliable models.


An important motivation for the inference of stochastic models from limited, non-ergodic MD data is the attempt to predict kinetic properties without passing through cumbersome free-energy calculation techniques, often based on biased simulations~\cite{pietrucci17,chipot23}.
We test the quantitative predictive power of the data-driven models of the fullerene dimer in water by comparing their dissociation rates with the brute-force one: as shown in Table 1, the model with the best performance from the viewpoint of diagnostics ($\tau=0.5$~ps) overestimates the MD rate by only 20\%, whereas models with different $\tau$ have less accurate rates, but always in the same order of magnitude. Note that without our correction scheme, the errors in the estimation of the barrier (and friction) lead to very severe errors in the rates, that sometimes cannot even be computed.


\begin{table}
\begin{tabular}{l|l}
$\tau$ [ps] &  MFPT [ns]\\ \hline
0.1        & $1.3 \pm 0.1$      \\
0.2        & $1.6 \pm 0.2$      \\
0.3       & $2.7 \pm 0.3$      \\
0.4        & $2.8 \pm 0.3$      \\
0.5        & $4.3 \pm 0.4$      \\
1        & $2.3 \pm 0.2$     \\ \hline
MD         & $3.6 \pm 0.2$
\end{tabular}
\caption{Mean first passage time (MFPT) for fullerene dissociation predicted by the Langevin models, inferred using the likelihood correction, computed using Euler-Maruyama integrator, compared to the brute-force MFPT from all-atom MD simulations. Statistical uncertainties are standard deviations of the mean.}
\end{table}



In summary, the results presented in this paper demonstrate that the proposed correction to likelihood inference 
eliminates systematic errors originating from finite-difference CV velocity estimation. The new scheme is able to parametrize accurate underdamped Langevin models of real complex systems projected on a CV, even for barrier-crossing processes, despite of the use of short, non-ergodic trajectories.
We underline the importance of the diagnostic tests we put forward to assess the pertinence and expected accuracy of the model in realistic scenarios, where the brute-force free-energy landscape or kinetic rate are not known beforehand. The main limitation to the applicability of this method, based on the Markovian Langevin equation, arises from the existence of long-time memory kernels in several realistic systems. To go beyond this hurdle, two possible strategies could be explored: i) increasing the dimensionality of the model to mitigate non-Markovian effects, or ii) adapting our approach 
to the inference of generalized Langevin models, where memory effects are reproduced by the addition of hidden variables, based on expectation-maximization likelihood schemes~\cite{Vroylandt22}.

\section{Acknowledgements}

This work was performed with the support of the Institut des Sciences du Calcul et des Données (ISCD) of Sorbonne University (IDEX SUPER 11-IDEX-0004).
Calculations were performed on the GENCI-IDRIS French national supercomputing
facility, under grant numbers A0110811069, A0130811069.
We gratefully acknowledge
Line Mouaffac and Jeremy Diharce for suggestions related to code development and simulations.


\bibliographystyle{apsrev4-2}
%
\newpage
\end{document}


\title{{\sl Supporting Information:\\
{\bf Correcting systematic parametrization errors in  underdamped Langevin models of molecular dynamics trajectories}}\\
}
\author{David Daniel Girardier}
\email{david.girardier@sorbonne-universite.fr}
\affiliation{Sorbonne Université, Musée National d’Histoire Naturelle, UMR CNRS 7590, Institut de
Minéralogie, de Physique des Materiaux et de Cosmochimie, IMPMC, F-75005 Paris, France}

\author{Hadrien Vroylandt}
\affiliation{Université Caen Normandie, ENSICAEN, CNRS, Normandie Univ, GREYC UMR 6072, F-14000 Caen, France}

\author{Sara Bonella}
\affiliation{Centre Européen de Calcul Atomique et Moléculaire (CECAM), Ecole Polytechnique Fédérale
de Lausanne, Lausanne 1015, Switzerland}

\author{Fabio Pietrucci}
\affiliation{Sorbonne Université, Musée National d’Histoire Naturelle, UMR CNRS 7590, Institut de
Minéralogie, de Physique des Materiaux et de Cosmochimie, IMPMC, F-75005 Paris, France}

\date{\today}

\maketitle


\subsection{Cumulants entering the propagator expression}

We list here the cumulants obtained with the procedure from Ref.~\cite{Drozdov97}
\begin{equation}
    \langle q\rangle = q_n + v_n\tau + \frac{1}{2}(\phi_n - \gamma v_n) \tau^2 
\end{equation}
\begin{equation}
    \langle v\rangle = v_n + (\phi_n -\gamma v_n )\tau - \frac{1}{2}[ (\phi_n -\gamma v_n) \gamma - \phi_n^q v_n] \tau^2
\end{equation}
\begin{equation}\label{eq:Mqq}
    M_{qq} = \langle q^2 \rangle -  \langle q \rangle  \langle q \rangle =  \frac{2}{3}c\tau^3 
\end{equation}
\begin{align}\label{eq:Mvv}
\begin{split}
    M_{vv} =& \langle v^2 \rangle -  \langle v \rangle  \langle v \rangle =  2c\tau - 2c\gamma\tau^2 + \frac{1}{3} (2c\phi_n^q + 4c\gamma^2)\tau^3 \\
\end{split}
\end{align}
\begin{equation}\label{eq:Mqv}
    M_{qv} = \langle qv \rangle -  \langle q \rangle  \langle v \rangle =  c\tau^2 - c\gamma\tau^3
\end{equation}

with $c=k_BT \gamma /m$ and $\phi_n^q$ is the positional derivative of the force at $q_n$.
The covariance matrix can then be written as
\begin{equation}\label{eq:detM}
    M =
    \begin{bmatrix}
        M_{qq} & M_{qv} \\
        M_{qv} & M_{vv} 
\end{bmatrix}
\end{equation}

\subsection{Correspondence between the propagator and the integrator}

In order to demonstrate the correspondence (up to order $\tau^2$) between the propagator derived from Ref. \cite{Drozdov97} and the VEC integrator~\cite{Vanden2006}, we start by rewriting the integrator in vector form:

\begin{equation}
    \begin{pmatrix}
        q_{n+1}  \\    v_{n+1}
    \end{pmatrix}
     = \begin{pmatrix}
        q_n\\    v_n 
     \end{pmatrix}+ \begin{pmatrix}
        \tau -  \f{\gamma\tau^2}{2}\\    -\gamma\tau  +\f{\gamma^2\tau^2}{2}
     \end{pmatrix} v_{n} 
    +\f{\tau}{2}\begin{pmatrix}
          \phi_{n}\tau \\ \left( 1-\gamma\tau\right)\phi_{n} + \phi_{n+1}
    \end{pmatrix}
    +\begin{pmatrix}
        \f{\tau}{2\sqrt{3}}\sigma & \f{\tau}{2}\sigma\\ - \f{\gamma\tau}{2\sqrt{3}}\sigma & (1 - \f{\gamma\tau}{2})\sigma
    \end{pmatrix} \begin{pmatrix}
     \tilde{G}_{n}\\  G_{n}
    \end{pmatrix}  
\end{equation}
 
We need to estimate the average of the variables
$q_{n+1}$ and $v_{n+1}$ of the Gaussian in Eq.~3 of the main text, therefore we evaluate the average of the right hand side of Eq.~S7, keeping $q_n$ and $v_n$ as constant (since the relevant dynamical variables are the increments $q_{n+1}-q_n$ and $v_{n+1}-v_n$).
The resulting moments $\langle q \rangle$, $\langle v \rangle$ are manifestly equivalent to the ones from the previous section, as long as we approximate $\phi_n^q v_n
\equiv \left(\frac{d\phi}{dq}\right)_{q=q_n}
\left(\frac{dq}{dt}\right)_{q=q_n}
\approx \f{\phi_{n+1}-\phi_{n}}{\tau}$. The covariance matrix is 
$\Tilde{M} = \Sigma\Sigma^T$ 
where $\Sigma$ is the matrix multiplying the noises in Eq.~S7. Explicitly,

\begin{equation}\label{eq:Mqq2}
    \tilde{M}_{qq} =   \frac{2}{3}c\tau^3 
\end{equation}
\begin{align}\label{eq:Mvv2}
\begin{split}
    \tilde{M}_{vv} =&   2c\tau - 2c\gamma\tau^2 + \frac{2}{3} c\gamma^2 \tau^3 \\
\end{split}
\end{align}
\begin{equation}\label{eq:Mqv2}
    \tilde{M}_{qv} =  c\tau^2 - \f{2}{3}c\gamma\tau^3
\end{equation}
Clearly, $\tilde{M}_{vv}$ and $\tilde{M}_{qv}$ only differ from ${M}_{vv}$ and ${M}_{qv}$ (see previous section) at the order $\tau^3$.

\subsection{Finite-difference definition of velocities}

Starting from the following definition of finite-difference velocities: 
\begin{equation}
    u^\lambda_{n} = (1-\lambda)\f{ q_{n}- q_{n-1}}{\tau} +\lambda \f{ q_{n+1}- q_{n}}{\tau}
\end{equation}
we can express the velocity for the VEC integrator as 
\begin{multline}
  u^\lambda_{n} =  \left( 1 -  \f{\gamma\tau}{2}\right) [\lambda v_{n} + (1-\lambda)v_{n-1}] +  \f{\tau}{2}[\lambda \phi_n + (1-\lambda)\phi_{n-1}  ] \\+ \f{\sigma}{2}    \lambda [G_{n} + \f{1}{\sqrt{3}}\tilde{G}_{n}] + \f{\sigma}{2}    (1-\lambda )[G_{n-1} + \f{1}{\sqrt{3}}\tilde{G}_{n-1}]
\end{multline}
Using the update equation for the velocity, this leads to
\begin{multline}
  u^\lambda_{n} =   \left( 1 -  \f{\gamma\tau}{2}\right) \f{1-\lambda\left( \gamma\tau - \f{\gamma^2\tau^2}{2}\right)}{\alpha} v_{n} 
  +  \f{\tau}{2}[\lambda - \f{(1-\lambda) \left( 1 -  \f{\gamma\tau}{2}\right)}{\alpha}] \phi_n \\
   + \f{\tau}{2}[(1-\lambda) - \f{(1-\lambda) \left( 1 +  \f{\gamma\tau}{2} + \f{\gamma^2\tau^2}{2}\right) }{\alpha}] \phi_{n-1} \\ 
 +  \f{\sigma}{2}    \lambda [G_{n} + \f{1}{\sqrt{3}}\tilde{G}_{n}]
 -\f{\sigma}{2}    \f{(1-\lambda )}{\alpha}[1-\gamma\tau] G_{n-1}
 + \f{\sigma}{2}    \f{(1-\lambda )}{\alpha} \f{1}{\sqrt{3}}\tilde{G}_{n-1} \
\end{multline}
with
\begin{equation}
    \alpha= \left(1-\gamma\tau  +\f{\gamma^2\tau^2}{2}\right).
\end{equation}
The Taylor expansion up to order  $\tau$ gives
\begin{multline}
  u^\lambda_{n} =  (1+ \gamma\tau(1/2-\lambda)) v_{n}   +  \f{\tau}{2}(1-2\lambda )\phi_{n}+  \f{\sigma}{2}    \lambda [G_{n} + \f{1}{\sqrt{3}}\tilde{G}_{n}] \\ 
  -\f{\sigma}{2}    (1-\lambda ) G_{n-1} + \f{\sigma}{2}    (1-\lambda ) \f{1+ \gamma\tau}{\sqrt{3}}\tilde{G}_{n-1}
\end{multline}
Therefore, the difference is minimized for $\lambda = 1/2$. 

\begin{equation}
  u^{1/2}_{n} = v_{n}   +  \f{\sigma}{4} \left[G_{n} + \f{1}{\sqrt{3}}\tilde{G}_{n} - G_{n-1}+\f{1}{\sqrt{3}}\tilde{G}_{n-1}\right] 
\end{equation}

Even though the derivation of this correction does not explicitly account for position-dependent friction. If we consider a case, where $\tau$ is small and $\gamma(q)$ do not vary drastically, it seems reasonable to allow the friction to be position dependent in the likelihood maximization.
\subsection{Correlations of position, velocity and noise for the VEC integrator}
In order to express the differences in Eq.~6, we will need the following properties for the noises times velocity 
\begin{align*}
  &\langle v_n G_n\rangle = 0
  &\langle v_n \tilde{G}_n\rangle= 0\\
  &\langle v_n G_{n-1}\rangle = \sigma -\gamma \frac{\tau}{2}\sigma
  &\langle v_n \Tilde{G}_{n-1}\rangle = -\gamma \frac{\tau}{2\sqrt{3}}\sigma\\
  &\langle v_n G_{n-2}\rangle = \sigma -\gamma \frac{3\tau}{2}\sigma + \mathcal{O}(\tau ^2)
  &\langle v_n \Tilde{G}_{n-2}\rangle = -\gamma \frac{\tau}{2\sqrt{3}}\sigma + \mathcal{O}(\tau ^2)
\end{align*}
and noises times position
\begin{align*}
  &\langle q_n G_n\rangle = 0
  &\langle q_n \tilde{G}_n\rangle= 0\\
  &\langle q_n G_{n-1}\rangle = \frac{\tau}{2}\sigma
  &\langle q_n \Tilde{G}_{n-1}\rangle = \frac{\tau}{2\sqrt{3}}\sigma\\
  &\langle q_n G_{n-2}\rangle = \frac{3\tau}{2}\sigma + \mathcal{O}(\tau ^2)
  &\langle q_n \Tilde{G}_{n-2}\rangle = \frac{\tau}{2\sqrt{3}}\sigma + \mathcal{O}(\tau ^2)
\end{align*}

All of these properties can be obtained by multiplying Eq.~4 by a certain noise $G_n$ or $\tilde{G}_n$ and taking the averages.

\subsection{Likelihood correction}
We show here how to isolate the correlations introduced in Eq.~6 starting from the likelihood function. For the sake of simplicity we use the following notations $q'\equiv q_{n+1}$,$q\equiv q_{n}$,$v'\equiv v_{n+1}$ and $v\equiv v_{n}$. Let us consider
\begin{align}\label{eq:logL}
    -\log\mathcal{L}_\theta =& \frac{1}{N}\sum^N \frac{1}{2}\textrm{ln}(4\pi \textrm{detM})  - \frac{1}{N}\sum^N \frac{M_{qq}}{2\textrm{det}M} (v' - M_v)^2  \\
    \nonumber&- \frac{1}{N}\sum^N \frac{M_{vv}}{2\textrm{det}M} (q' - M_q)^2 +  \sum^N \frac{1}{N} \frac{M_{qv}}{\textrm{det}M}(v' - M_v)(q'- M_q)   
\end{align}
We can take variances (Eq.~\ref{eq:Mqq},~\ref{eq:Mvv},~\ref{eq:Mqv}) out of the sums in Eq.~\ref{eq:logL} to express it as
\begin{align}\label{eq:logL2}
    -\log\mathcal{L}_\theta =& \frac{1}{N}\sum^N \frac{1}{2}\textrm{ln}(4\pi \textrm{detM})  - \frac{M_{qq}}{2\textrm{det}M} \langle (v' - M_v)^2 \rangle \\
    \nonumber&- \frac{M_{vv}}{2\textrm{det}M} \langle (q' - M_q)^2 \rangle+ \frac{M_{qv}}{\textrm{det}M} \langle (v' - M_v)(q'- M_q)\rangle
\end{align}
with 
\begin{align}
     \langle(q' - M_q)^2\rangle =& \langle q'^2\rangle  + \langle q^2\rangle  - 2\langle qq'\rangle  + \langle vq'\rangle (-2\tau + \gamma \tau^2) + \langle vq\rangle (2\tau - \gamma\tau^2) \\
     \nonumber & + \langle v^2\rangle (\tau^2 -\gamma\tau^3 + \frac{1}{4} \gamma^2\tau^4)
     - \langle \phi q'\rangle \tau^2 + \langle \phi q\rangle \tau^2\\
     \nonumber & + \langle \phi v\rangle (\tau^3 - \frac{1}{2} \gamma\tau^4) + \frac{1}{4}\langle \phi^2\rangle \tau^4
\end{align}

\begin{align}
     \langle (v' - M_v)^2\rangle  =& \langle v'^2\rangle  + \langle v'v\rangle (-2 + 2\gamma \tau - \gamma^2\tau^2) \\
     \nonumber&+ \langle v^2 \rangle (1-2\gamma \tau + 2\gamma^2\tau^2 - \gamma^3\tau^3 + \frac{1}{4}\gamma^4\tau^4) + \langle \phi v'\rangle(-2 + \gamma \tau^2) \\
     \nonumber& + \langle \phi v\rangle(2 -3 \gamma \tau^2 + 2 \gamma^2 \tau^3 - \frac{1}{2} \gamma^3 \tau^4) + \langle \phi^2\rangle(\tau^2 - \gamma\tau^3 + \frac{1}{4} \gamma^2 \tau^4)\\
     \nonumber& - \langle vv' \phi_q \rangle \tau^2 + \langle v^2 \phi_q \rangle \tau^2 + \langle v \phi\phi_q\rangle \tau^3 - \langle v \phi_q\rangle\gamma\tau^3-\f{1}{2} \langle v \phi\phi_q\rangle\gamma\tau^4 \\\nonumber& +\f{1}{2} \langle v^2\phi_q\rangle \gamma^2\tau^4 + \f{1}{4} \langle v^2\phi_q^2\rangle\tau^4
\end{align}

\begin{align}
     \langle (q' - M_q) (v' - M_v)\rangle  = &\langle v'q'\rangle  - \langle v'q\rangle     \\
     \nonumber & + \langle vq'\rangle(-1 + \gamma\tau - \frac{1}{2}\gamma^2\tau^2) +\langle vq\rangle (1-\gamma\tau + \frac{1}{2} \gamma^2\tau^2) \\
     \nonumber&+ \langle vv'\rangle (-\tau + \frac{1}{2}\gamma\tau^2)
      + \langle v^2\rangle ( \tau -\frac{3}{2}\gamma\tau^2 + \gamma^2\tau^3 - \frac{1}{4} \gamma^3 \tau^4)\\
      \nonumber& + \langle \phi q' \rangle (-\tau +\frac{1}{2} \gamma \tau^2 ) + \langle \phi q \rangle (\tau -\frac{1}{2} \gamma \tau^2 )\\
      \nonumber& - \frac{1}{2} \langle \phi v' \rangle + \langle \phi v \rangle (\frac{3}{2}\tau^2 - \frac{3}{2} \gamma \tau^3 + \frac{1}{2} \gamma^2 \tau^4) + \langle \phi^2\rangle (\frac{1}{2} \tau^3 - \frac{1}{4} \gamma\tau^4)\\ \nonumber& -\f{1}{2}\langle vq'\phi_q\rangle\tau^2 + \f{1}{2}\langle vq\phi_q\rangle\tau^2 + \f{1}{2}\langle v^2\phi_q\rangle\tau^3 \\
      \nonumber& + \f{1}{4}\langle v\phi\phi_q\rangle\tau^4 - \f{1}{4}\langle v^2\phi_q\rangle\gamma\tau^4
\end{align}
Now that the correlations of Eq.~6 appear in the previous expressions it is then possible to account for the use of finite difference velocities.



\subsection{Noise Analysis}

To further validate the accuracy of the inference, it is useful to analyze the noise that would have been required to reproduce the initial trajectories using the optimal Langevin model. In the absence of the issues arising from finite-difference velocities, the recovered noise should be centered around zero, exhibit unit variance, and be delta-correlated in time.
We focus on the positional part of the numerical integrator (Eq.~4), combining the two noise terms and isolating them as:
\begin{equation}
    \label{eq:noise1}
    \mathcal{G}_n = G_n + \frac{1}{\sqrt{3}}\tilde{G}_n = \Bigg( \frac{q_{n+1}-q_n}{\tau} - \Big(1-\frac{\gamma\tau}{2}\Big)v_n -\frac{\tau}{2} \phi_n\Bigg) \frac{2}{\sigma}
\end{equation}
Using properties of Gaussian white noises and Eq.~6, it is possible to obtain the following properties of $\mathcal{G}_n$:
\begin{equation}\label{eq:Grecovered}
    \langle \mathcal{G}_n\rangle = 0\ ,\  \langle \mathcal{G}_n^{2}\rangle = \frac{2}{3}\ ,\  \frac{\langle \mathcal{G}_n \mathcal{G}_{n+1}\rangle}{\langle \mathcal{G}_n^{2}\rangle} = \frac{1}{4} 
\end{equation}

while the time correlation vanishes for time delays larger than $\tau$.

\subsection{Accuracy of inferred friction coefficients}

Table~\ref{tab:friction} displays friction coefficient for the different time resolution in the underdamped Langevin benchmark case. We can observe that without using the correcting scheme the value obtained is $\approx 2/3\gamma$ (coherent with Ref.~\cite{ferretti20}). In this case, due to the non-ergodicity of the trajectories the error on the friction also impacts the inference of the free energy.

\begin{figure*}%
    \centering
    {\includegraphics[width=0.8 \linewidth]{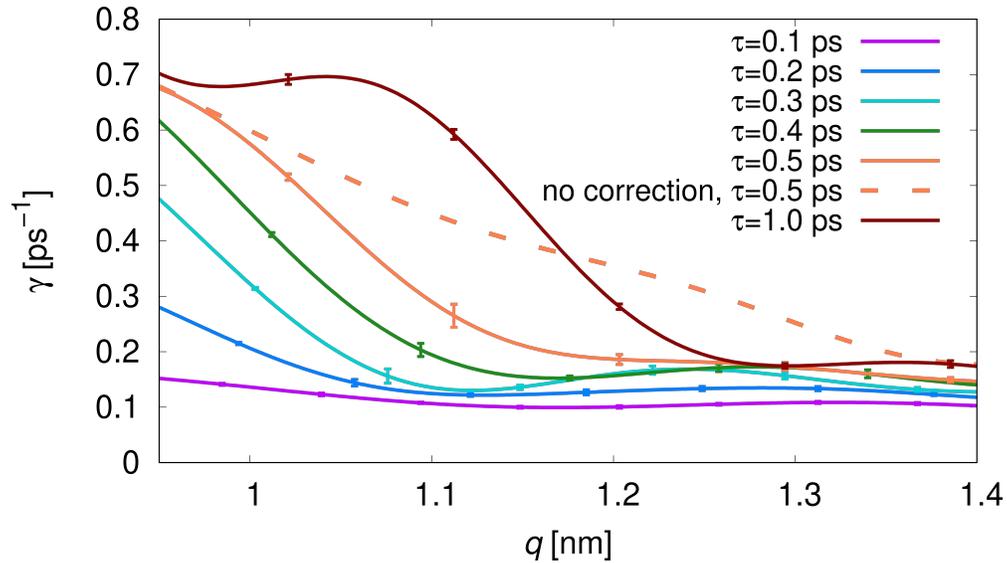} }%
    \caption{Friction profiles for the different Langevin models, for fullerene dimer dynamics in water, at different time resolutions $\tau$ using the likelihood correction (the dashed orange curve shows the result without correction).}%
    \label{fig:frictionFull}%
\end{figure*}

We also show the friction profiles for the fullerenes dimer in water (with modified mass) on Fig.~\ref{fig:frictionFull}.
\begin{table}
\label{tab:friction}
\begin{tabular}{l|l}
$\tau$ [ps] &  $\gamma$ [ps$^{-1}$]\\ \hline
0.001       & $4.92 \pm 0.01$   \\
0.001 (no corr.)     & $2.97 \pm 0.01$      \\
0.005        & $4.73 \pm 0.01$      \\
0.01        & $4.50 \pm 0.02$      \\
0.02      & $4.17 \pm 0.02$      \\
0.04        & $3.81 \pm 0.01$      \\
expected         & $5.0$
\end{tabular}
\caption{Friction for the different Langevin models, inferred using the likelihood correction, for the benchmark double-well system: underdamped Langevin trajectories relaxing from the top of a barrier, integrated with a 0.1~fs time step using an Euler-Maruyama integrator with unit mass and $\gamma = 5$~ps$^{-2}$. The case where the likelihood correction is not applied is reported as well.}
\end{table}
\newpage

\subsection{Fullerenes dimer in water with unmodified masses}

\begin{figure*}%
    \centering
    {\includegraphics[width=0.8 \linewidth]{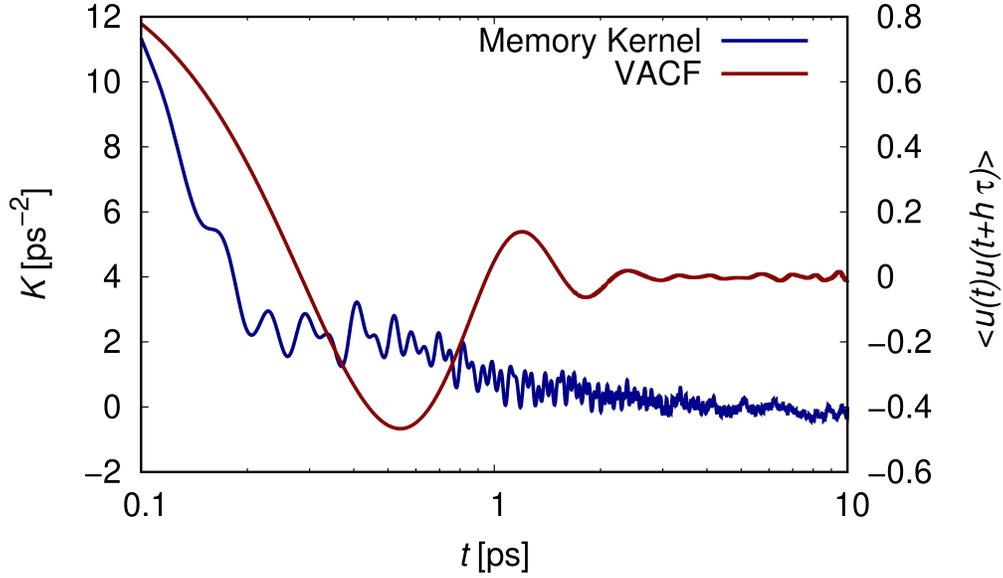} }%
    \caption{Memory kernel and VAC computed for 10ns in the associative bassin, for unmodified masses system.}%
    \label{fig:KVAC}%
\end{figure*}

Fig.~\ref{fig:KVAC} presents the memory kernel and velocity autocorrelation function (VACF) for a fullerene dimer in water, projected onto the distance between the two centers of mass. The masses of the carbon and oxygen atoms remain unmodified. We observe no clear timescale separation, indicating that the system does not exhibit an underdamped regime.

\begin{figure*}%
    \centering
    {\includegraphics[width=0.8 \linewidth]{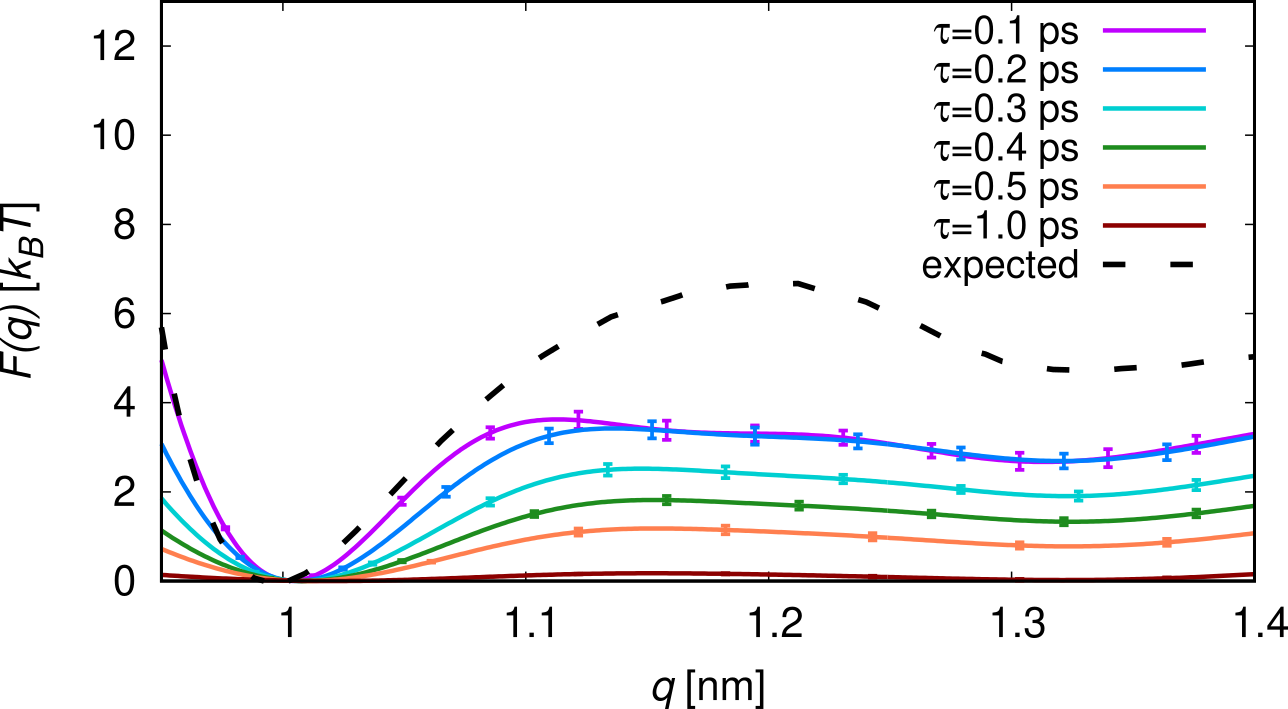} }%
    \caption{Free-energy profiles at different time resolutions $\tau$ using the likelihood correction, for unmodified masses system.}%
    \label{fig:Funmod}%
\end{figure*}

This lack of separation explains why none of the free-energy profiles inferred from the underdamped models agree with the expected free energy (see Fig~\ref{fig:Funmod}).


\bibliographystyle{apsrev4-2}